\documentclass[%
 aip,
 amsmath,amssymb,
 reprint,%
]{revtex4-1}

\usepackage{graphicx}
\usepackage{dcolumn}
\usepackage{bm}

\usepackage{mathscinet} 
\usepackage{hhline}
\usepackage{enumerate}  
\usepackage[utf8]{inputenc}
\usepackage[T1]{fontenc}
\usepackage{mathptmx}
\usepackage{etoolbox}
\usepackage{makecell}

\usepackage{calrsfs}
\DeclareMathAlphabet{\pazocal}{OMS}{zplm}{m}{n}
\SetMathAlphabet\pazocal{bold}{OMS}{zplm}{bx}{n}

\newtheorem{theorem}{Theorem}

\newtheorem{lemma}[theorem]{Lemma}

\makeatletter
\def\@email#1#2{%
 \endgroup
 \patchcmd{\titleblock@produce}
  {\frontmatter@RRAPformat}
  {\frontmatter@RRAPformat{\produce@RRAP{*#1\href{mailto:#2}{#2}}}\frontmatter@RRAPformat}
  {}{}
}%
\makeatother
\begin{document}


\title[No-slip Billiards with Particles of Variable Mass Distribution]{No-slip Billiards with Particles of Variable Mass Distribution}
\author{J. Ahmed}
\affiliation{Department of Mathematics, University of Delaware, Ewing Hall, Newark, DE 19711 }
\author{C. Cox}%
\affiliation{Department of Mathematics, University of Delaware, Ewing Hall, Newark, DE 19711 }

\email{clcox@udel.edu.}

\author{B. Wang}
\affiliation{Department of Mathematics, University of Delaware, Ewing Hall, Newark, DE 19711 }

\date{\today}

\begin{abstract}
Astute variations in the geometry of mathematical billiard tables have been and continue to be a source of understanding their wide range of dynamical behaviors, from regular to chaotic. Viewing standard specular billiards in the broader setting of no-slip (or rough) collisions, we show that an equally rich spectrum of dynamics can be called forth by varying the mass distribution of the colliding particle.
We look at three two-parameter families of billiards varying both the geometry of the table and 
the particle, including as special cases examples of standard billiards demonstrating dynamics from integrable to chaotic, and show that markedly divergent dynamics  may arise by changing only the mass distribution. 
Furthermore, for certain parameters billiards emerge which display unusual dynamics, including examples of full measure periodic billiards, conjectured to be nonexistent for the standard billiards in Euclidean domains. 

\end{abstract}

\maketitle

\begin{figure}[htbp]
\includegraphics[width=.48\textwidth]{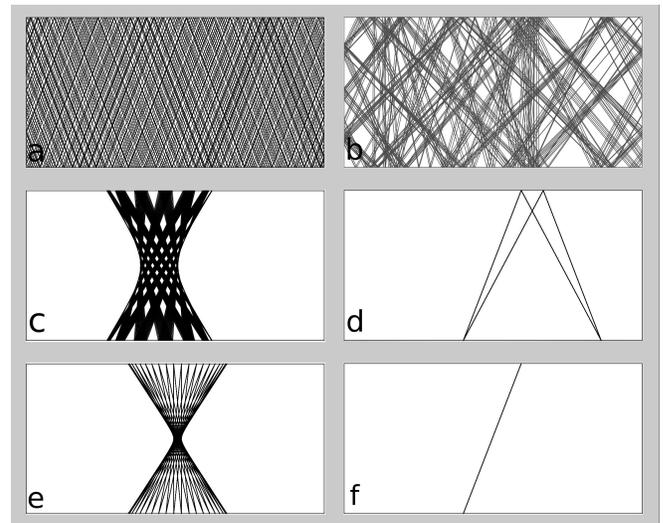}

\caption{\small A single orbit in a rectangular no-slip billiard table, with identical initial positions, linear and rotational velocities, and number of collisions, varying only the particle's radially symmetric mass distribution. 
In a) $(\eta=0)$ all the mass is at the center, dynamically equivalent to the classical billiard model; in b) $(\eta=.05)$ the mass is distributed slightly outward and the regularity is marred; c) $(\eta_u \approx .39)$ shows uniform mass distribution; 
in d) $(\eta=\frac{1}{2})$ the particles are ring shaped and the billiard $4$-reflective; e) and f) $(\eta=.95,1)$ show  systems with mass distributed beyond the radius of collision. }
\label{fig:etavary} 
\end{figure}

The standard specular billiard model in dimension two has proven widely useful, both as a  model for an extensive variety of applications and as a tool for better understanding dynamical systems in general. 
Many of the insights in the latter case have been facilitated by the fundamental link between the geometry of the billiard table and the dynamics, giving a flexible tool for exploration. 
Most commonly, the colliding particle is viewed as a point mass, simplifying the model but also ceding an avenue for controlling the dynamics of the system. 
However, new directions emerge if one starts from a more physically motivated model in which the colliding particles have a positive radius and a known (usually non-uniform) mass distribution.  
We consider disk-shaped particles 
with radial symmetry, varying the systems by moving the mass inward or outward according to a mass distribution constant $0\leq \eta \leq 1$, defined explicitly in Section \ref{sec:no-slip}, with $\eta=0$ corresponding to the standard billiard model. 
Varying only this parameter  may yield distinctly different dynamics, as shown in the orbits resulting from identical starting trajectories on identical rectangular tables in Figure \ref{fig:etavary}. 

In this paper we follow this idea down two avenues. 
First, we numerically investigate three often studied examples of standard billiards in this broader setting. The well-known integrable ellipse billiard, a paragon of regular dynamics, is shown to retain some regions of regularity as $\eta$ increases, but they are surrounded by regions of less regular dynamics. 
We also look at a family of Sinai billiards and a family of billiards with three arc-shaped edges, including the three-petal flower. 
These families which include iconic examples  of dispersion and defocusing, the two mechanisms for producing ergodicity in standard billiards, evolve to produce  examples featuring elliptic islands and even stronger regular behavior when extended to include positive values of the mass distribution parameter $\eta$. 

The second avenue we pursue is showing that new categories of dynamic behavior unknown in the standard model may be found in this more general context. We give analytic and numeric results showing that periodic orbits are ubiquitous, and in particular show that Ivrii's Conjecture for standard billiards in a Euclidean domain, which states that there exist no \textit{k-reflective} billiards (having postive measure regions comprised of k-periodic orbits), fails to hold for no-slip billiards. In fact, we construct new examples of \textit{persistent periodicity}, billiards for which the set of periodic points is full measure.

\section{\label{sec:introduction}Introduction}

In recent years, many variations of the standard billiard model have garnered interest. Moving away from elementary rigid bodies motion the specular collision model may be replaced with stochastic boundaries,\cite{chumley} some form of a potential as in composite\cite{composite} billiards  or magnetic billiards,\cite{Datseris} or optically inspired refractions\cite{Davis, deblasi} 
to name a few. Even in the realm of simple, physically 
motivated particle collisions, natural refinements exist which have yet to be fully explored. For example, it was recently shown that for certain billiard tables in dimension two, whether the particle is a point mass or  has positive radius may determine whether the system exhibits regular or chaotic behavior, absent any deeper modifications to the model.\cite{Bun2} Here we look at some of the dynamics that may appear if a positive radius particle is assumed to collide with a rough surface with nondissipative friction according to the no-slip model. For the (generally convex) tables we consider there will be no complications arising merely from the nonzero radius, but the rotational inertia of the particle will induce significant variations in the dynamics.

No-slip billiards are dynamical systems in which rotating particles move freely between boundaries and then reflect according to the no-slip rigid body collision model, in which linear and angular momentum may be conservatively exchanged. 
Accordingly, the rotational inertia of the particle will influence the dynamics when it is non-zero.
The no-slip alternative naturally arises as a collision model for two freely moving bodies in $\mathbb{R}^n$ for any $n \geq 2$, along with the standard model, under assumptions of conservation of energy, (collective) conservation of linear and angular momentum, time reversability, and the rigid body condition that the force is applied at the point of contact.\cite{CF} 
Frequently, the two maps are viewed as frictionless and completely rough alternatives, assuming particles of uniform mass distribution, but this two-body collision model may with equal justification be viewed as yielding a continuous family of alternatives to the specular model, with the two possibilities converging if the mass is concentrated at a point.\footnote{The model arising as the small radius limit of the no-slip (or ``rough'') alternative gives a complete reversal of rotational velocity at collisions, while the specular case is generally thought to be frictionless and accordingly leaving rotation unchanged. Projecting out the rotation, however, yields identical models.} Billiard tables may be seen as arising 
by taking the limit as the mass of the first body is taken to infinity, rendering it fixed. From this perspective, much of the progress in understanding billiard dynamics arises from insightful choices for the geometry of the first body, that is, the table. It is natural to ask what may be gleaned from varying the second body, that is, the finite mass particle. 

The idea of investigating the dynamics of a billiard system by varying the colliding particle is not new. For example, efforts have been made to incorporate rotation within a Hamiltonian framework (for example using non-spherical particles\cite{cowan} or paired spheres in a dumbbell shape\cite{baryshnikov}) and addressing the added complexities of the asymmetry. One advantage of our model is that it retains the simplicity of calculation and numerical simulation afforded by a circular disk particle while at the same time accessing the versatility which arises by varying the particle geometry.    
Specifically, we will fix the mass and the radius, then consider a mass distribution\footnote{Since the no-slip model depends only on the moment of inertia, each $\eta$ actually corresponds to an equivalence class of mass distributions.} constant $0\leq \eta \leq 1$ which corresponds, for example, to a point mass at $\eta=0$, to a ring shaped particle with all the mass on the boundary when $\eta = \frac{1}{2}$, and to the mathematically reasonable though physically unattainable infinite moment when $\eta=1$. In particular, we allow configurations in which the mass may be located beyond the radius of collision. Certainly, such systems seem less applicable as simple systems of colliding particles, but there is no fundamental physical reason to exclude them. Mathematically, viewed abstractly as dynamical systems, they are quite natural, although the $\eta=1$ case reduces to triviality with every orbit $2$-periodic. However, even this physically dubious extreme is instructive, the pinnacle of a larger trend towards a prevalence of periodicity as $\eta$ increases. 

In Section \ref{sec:no-slip} we give a brief history of the no-slip model and a few results we will need, including a discussion of the phase space of no-slip billiards.
\begin{figure}[htbp]
    \centering

    \includegraphics[width=.48\textwidth]{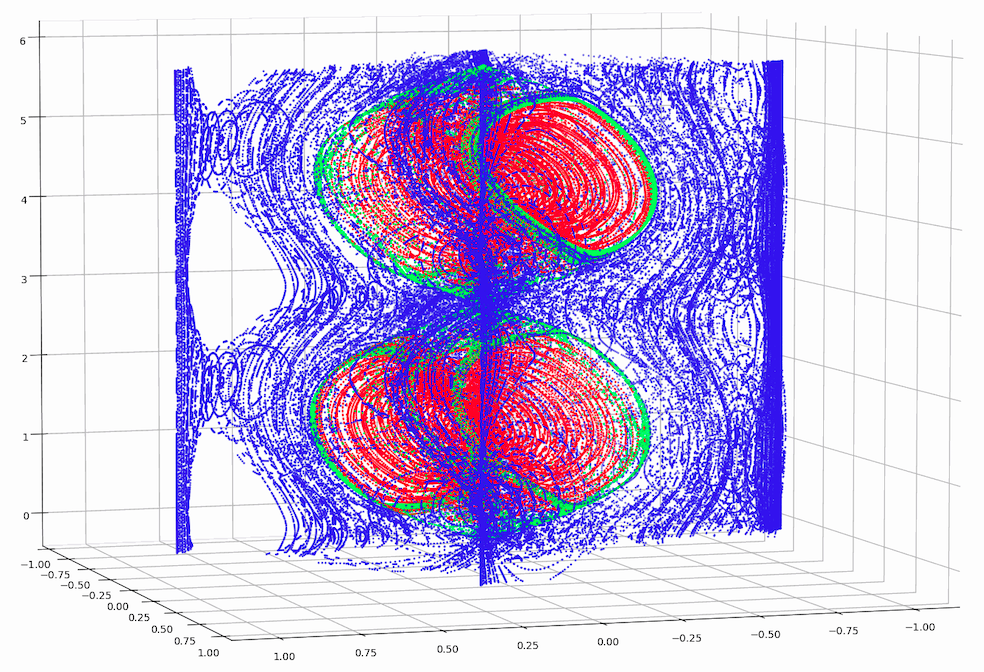}
    \caption{\small A 3D phase portrait of the uniform mass distribution no-slip ellipse $E(.3,\eta_u)$, with orbits sampled to reveal the inner structure. Orbits corresponding to quasiperiodic orbits in the elliptic region of the specular case (shown in red) seem to maintain an integrable structure. Outer orbits (blue) display a more complicated structure. Green orbits pass through the foci.   }    
\label{fig:3Dellipsephase}
\end{figure}
In Section \ref{sec:3billiards} we numerically investigate three well-known families from classical billiards, considering the dynamics when mass distribution of the particle is varied. 
Specifically, we look at three two-parameter families of billiards, with the first parameter varying the geometry of the table and the second parameter $\eta$.

The first we denote by
$S(R,\eta)$, Sinai type billiards on a unit torus with a dispersing disk of diameter $R$ and collisions given by the no-slip model with mass distribution constant $\eta$. Only $S(R,0)$, corresponding to a standard dispersing billiard for any $R$, is ergodic. However, there is apparently a positive measure ergodic component for other cases, with its measure varying inversely to both $R$ and $\eta$. 

Secondly, we consider the two parameter family $E(e,\eta)$, where $e$ is the eccentricity of the ellipse boundary. 
The integrability of the central region of phase space surrounding the elliptic periodic point in the standard $E(e,0)$ billiards persists for particles with positive inertia, but more complicated dynamics develop in the outer regions for the cases $E(e,\eta)$, $\eta>0$. Figure \ref{fig:3Dellipsephase} shows a phase portrait (in three dimensions with the rotational dimension) for the uniform mass case, $E(.3, \eta_u)$, where $\eta_u= \frac{1}{\pi}\arccos{\frac{1}{3}}\approx .392$. 

The third family $F(\theta,\eta)$, with $-\frac{\pi}{6}< \theta < \frac{2\pi}{3}$, consists of billiards with vertices on an equilateral triangle and edges which are arcs of circles, all three of equal curvature, parametrized by the angle relative to the straight segment. Hence, $F(0,\eta)$ consists of no-slip equilateral triangles, while any $\theta<0$ yields a dispersing billiard (for the standard case) and sufficiently large $\theta$ will correspond to the three petal flower, known in the standard case to be ergodic through the defocusing mechanism. The varied dynamics demonstrated in the specular one parameter subfamily $F(\theta,0)$ spawns yet greater variety in the generalized setting. The equilateral triangle $F(0,\eta)$, for a generic $\eta>0$, appears to have a phase space comprised entirely of periodic points and quasiperiodic points in surrounding elliptic islands. For certainly parameters, however, the regularity is even stronger. In particular, the uniform mass distribution equilateral triangle $F(0, \eta_u)$ is known to be persistently periodic, with 4-reflective and 6-reflective sets combining to have fully measure.\cite{CFII}
\begin{figure}[htbp]
    \centering
    \includegraphics[width=.48\textwidth]{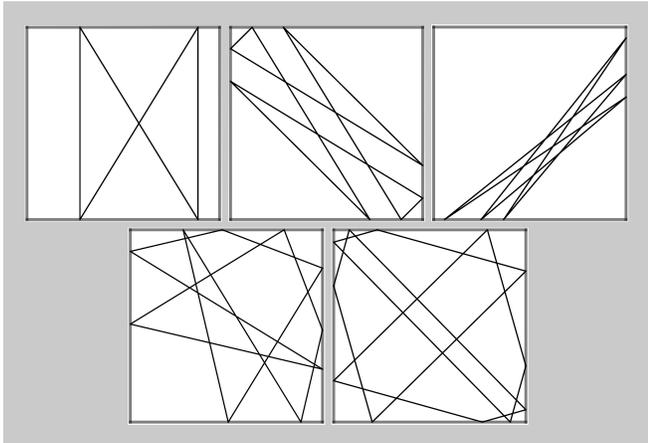}
    \caption{\small In sharp distinction to specular billiards, orbits in the no-slip square billiard with ring-shaped particles. 
    having mass concentrated on the rim, will be periodic with one of the five types above  regardless of the initial position, spin, or velocity.}    
\label{fig:PPSquare}
\end{figure}
The prevalence of periodic orbits and examples of this stronger regularity are the subjects of Section \ref{sec:pp}. While non-generic, persistent periodicity in no-slip billiards is nonetheless commonly occurring. For open tables such as infinite strips bounded by a pair of horizontal lines or wedges of any angle less than $\pi$, it is known that for $\eta>0$ the portion of phase space corresponding to non-escaping orbits has positive measure. Reducing to the phase space of non-escaping orbits,  the following theorem holds.
\begin{theorem}
Values for the particle mass distribution constant $0 \leq \eta \leq 1$ yielding persistently periodic billiards are dense.  In particular, for any integer $n>1$,
\begin{enumerate}[(1)]
\item  there exists an $\eta$ so that all points in phase space of the infinite strip are $2n$-periodic, and 
 \item for any angle $0 < \phi
\leq \frac{\pi}{8}$, there is a mass distribution constant $\eta$ which makes the wedge of angle $2\phi$ persistently periodic of period $2n$.
 \end{enumerate}
 \label{thm:main}
\end{theorem}

Among closed billiards, we search for new persistently periodic examples in addition to the known uniform mass equilateral triangle, giving numerical evidence for several candidates, including the square with ring-shaped particles ($\eta=\frac{1}{2}$) in Figure \ref{fig:PPSquare} and a persistently periodic pentagon. We also show that dual examples (in the sense $\eta_2=1-\eta_1$) with mass beyond the radius of collision exist in the cases of the triangle and pentagon.

\section{\label{sec:no-slip}A brief review of the no-slip model}

Though the dynamics of general no-slip billiards are less well understood than those of the specular case (for example, it has not yet been analytically shown that a single example exists where the no-slip billiard map is ergodic) the model is natural and has arisen apparently independently from many directions of inquiry.  A three dimensional model was described by  Richard Garwin in 1969,\cite{garwin} referred to there and elsewhere in the physics literature as \textit{rough billiards}. Broomhead and Gutkin  derived the case of uniform mass disks in dimension two and first looked at the long term dynamics of such systems, showing the fundamental result that the no-slip strip was bounded for any non-parallel starting trajectory or spin,\cite{gutkin} a result generalized in Theorem \ref{thm:main}.  The model has been incorporated into a Lorentz gas to allow for a transport model with rotating scattering disks.\cite{MLL} More recently, the no-slip model was derived very generally, starting with a model for rigid collisions between two bodies described by a wide class of measurable sets in $\mathbb{R}^n$ for any $n\geq 2$, with the known models in dimensions two and three arising as special cases.\cite{CF}

\begin{figure}[htbp]
    \centering

    \includegraphics[width=.48\textwidth]{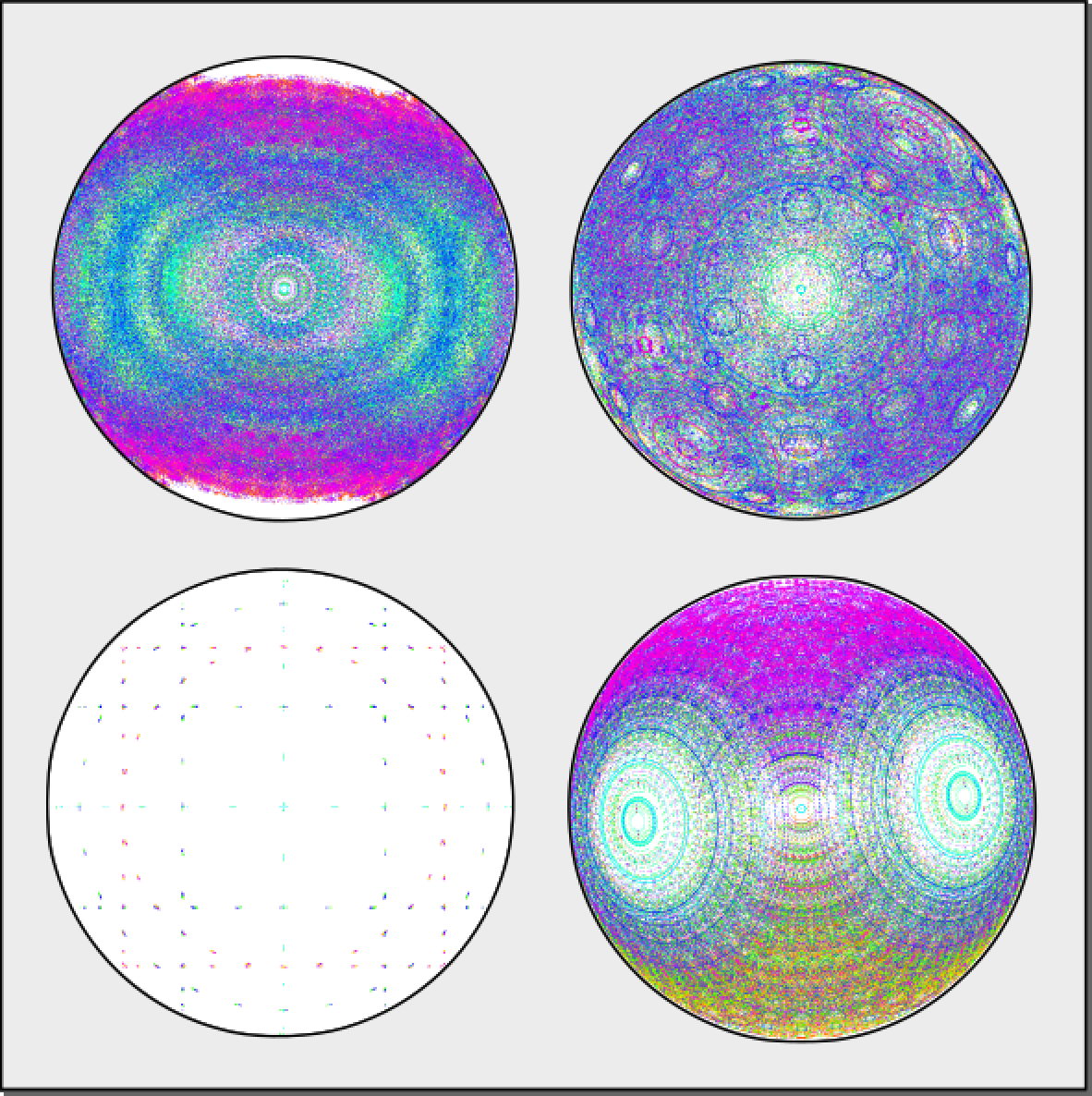}
    \caption{\small Velocity phase portraits, colors indicating distinct orbits, for four of the rectangular billiards from Figure \ref{fig:etavary}, illustrating the variation in dynamics from changing only the mass distribution of the disk-shaped particle. Here $\eta$ goes from $0.05$ (upper left), to uniform mass  $\eta_u \approx 0.392$ (upper right), to ring-shaped $0.5$ (lower left), and finally $0.95$ (lower right).  }    
\label{fig:rectvpp}
\end{figure}

It has long been known that standard billiards may be seen as arising as limits of geodesic flows through a flattening of the Riemannian manifold, \cite{Arnold} and more recently this idea was extended to \textit{rolling billiards} arising from considering nonholonomic systems.  No-slip billiards, or \textit{nonholonomic billiards}, on the infinite strip (or the ellipse) were derived from the flattening of a nonholonomic rolling system on a cylinder (or ellipsoid).\cite{BKM} It was subsequently shown that  the small radius limit of rolling systems yield no-slip billiards more generally.\cite{rolling}  While there is no explicit nonholonomic assumptions in the standard mechanical derivation of no-slip billiards, the link to nonholonomic systems can also be bridged from that direction, as it has been  shown that for systems of no-slip spheres in a cylinder under an external force, the small-bounce limit can approach the known motion of nonholonomic rolling.\cite{CCCF}

For no-slip billiards in the plane, with the added rotational dimension the billiard flow is \textit{a priori} a map on the six dimensional tangent bundle. 
However, the dynamics are encapsulated by the billiard map on a reduced three dimensional phase space, beginning with the standard simplifications of restricting to the unit tangent bundle and looking at the collision points on the boundary, parametrized by arclength.\cite{chernov}
The velocity vector, including rotational velocity, is then in the upper half sphere $S^{2+}$.  Additionally, we ignore the rotational position, which does not contribute to the dynamics generally,\footnote{An interesting exception to this, however, occurs in the case of a particle which is half rough and half smooth.\cite{CF}} and the phase space is reduced to $\pazocal{M}=S^1 \times S^{2+}$, which we render as a cylinder, with the billiard map $$\pazocal{F}:\pazocal{M}\rightarrow \pazocal{M}$$ preserving the billiard measure. 
Here the factor $\cos \phi d\phi$ in the specular billiard measure generalizes to $\cos \phi dA$ where $dA$ is the Euclidean measure on $S^{2+}$, and $\phi$ is the angle from the normal in both cases.

Two projections are sometimes useful for simplifying the picture while still capturing dynamic features. By the velocity phase portrait we mean the projection of $S^{2+}$ to a disk, ignoring the position, with the convention that the horizontal direction corresponds to the tangential component of velocity at collisions and the vertical direction to the rotational component. Figure \ref{fig:rectvpp} shows the velocity phase portraits for the billiards give in Figure \ref{fig:etavary}. We can also  project to standard (arclength-angle) phase space, ignoring the rotational velocity. Note that orbits, and in particular elliptic islands, may overlap in these projections. 

We conclude this section with several results that underlie subsequent calculations of the no-slip billiard map.\cite{CFZ} For a rotationally symmetric disk-shaped particle, define $\gamma:=\sqrt{\frac{I}{mR^2}}$, where the moment of inertia $I$ is proportional to the mass $m$ and radius $R$ squared, so that $\gamma$ is independent of both. We now formally define our mass distribution constant
\begin{equation}
    \eta:=\frac{1}{\pi} \cos^{-1} \left(
    \frac{1-\gamma^2}{1+\gamma^2}
    \right).
    \label{eq:eta}
\end{equation}
At a point of collision, we define the orthonormal frame $(e_1,e_2,e_3)$ where $e_2$ is the unit tangent, $e_3$ the outward unit normal, and the rotational component  $e_1=e_2 \times e_3$. Note that the units for the rotational direction so defined are scaled by $\gamma R$, normalizing so that the energy of a velocity vector $\textbf{v}=<v_1,v_2,v_3>^\intercal$ is given by the Euclidean norm. The no-slip transformation matrix $T$ on the incoming velocity vector $\textbf{v}^-$, so that the outgoing vector is $\textbf{v}^+=T\textbf{v}^-$, is then given by  
\begin{eqnarray}
\label{eq:T}
T =
\begin{pmatrix} - \cos\left( \pi \eta \right) & - \sin\left( \pi \eta \right) & 0
\\ - \sin\left( \pi \eta \right) & \cos\left( \pi \eta \right) & 0
\\ 0 & 0 & -1 \\
\end{pmatrix}
\end{eqnarray}
where $\eta$ is as defined in Equation \ref{eq:eta}.

We briefly mention two other coordinate systems, in which two useful facts about the dynamics of no-slip billiards are conveniently stated. For an open wedge billiard with angle $2\phi$, let $\hat{e}_2$ be in the direction of the bisector of the wedge, $\hat{e}_3$ the planar direction orthogonal to $\hat{e}_2$, and $\hat{e}_1=\hat{e}_2 \times \hat{e}_3$. In these coordinates, starting from any point on the wedge with velocity $\textbf{v}=<v_1,v_2,v_3>$, the orbit will be $2$-periodic if and only if $v_2=0$ and
\begin{equation}
    \frac{v_1}{v_3}=-\frac{\sin \phi}{\gamma}.
    \label{eq:2per}
\end{equation}
For polygonal tables these periodic points, which are Lyapunov stable, are ubiquitous. Indeed, since the collision only depends on the tangent of the curve at the point of contact, such $2$-periodic orbits are found in most tables. However, $2$-periodic orbits between two boundaries with positive curvature may not be linearly stable.\cite{W,CFZ} The stability threshold in the case where both edges have positive curvature $k$, the distance between collisions is $d$, and the osculating wedge with orthogonal bisector has angle $2\phi$ is given by the relation
\begin{equation}
    \frac{1}{2}kd=\frac{1-\cos^2(\pi\eta)\cos^2\phi}{\cos^2(\pi\eta)\cos\phi}.
    \label{eq:stabthresh}
\end{equation}

Equation \ref{eq:2per} may be derived by calculating the eigenvector of the return map, and for a third coordinate system we use spherical coordinates $(\theta, \psi, \rho)$  about the axis given by the eigenvector. In these coordinates, for a no-slip billiard in a wedge, the return map on the velocity after two collisions fixes $\rho$ (by energy conservation), fixes angle $\psi$ from the axis, and rotates about the axis (that is, increases $\theta$) by a constant determined only by the wedge angle $2\phi$ and the mass distribution. Specifically, 
\begin{equation}
\cos(\theta)=1-8\cos^2 \frac{\pi}{2}\eta \cos^2 \phi  +8\cos^4 \frac{\pi}{2}\eta \cos^4 \phi.  
    \label{eq:rotate}
\end{equation}

\begin{figure}[htbp]
    \centering
    \includegraphics[width=.48\textwidth]{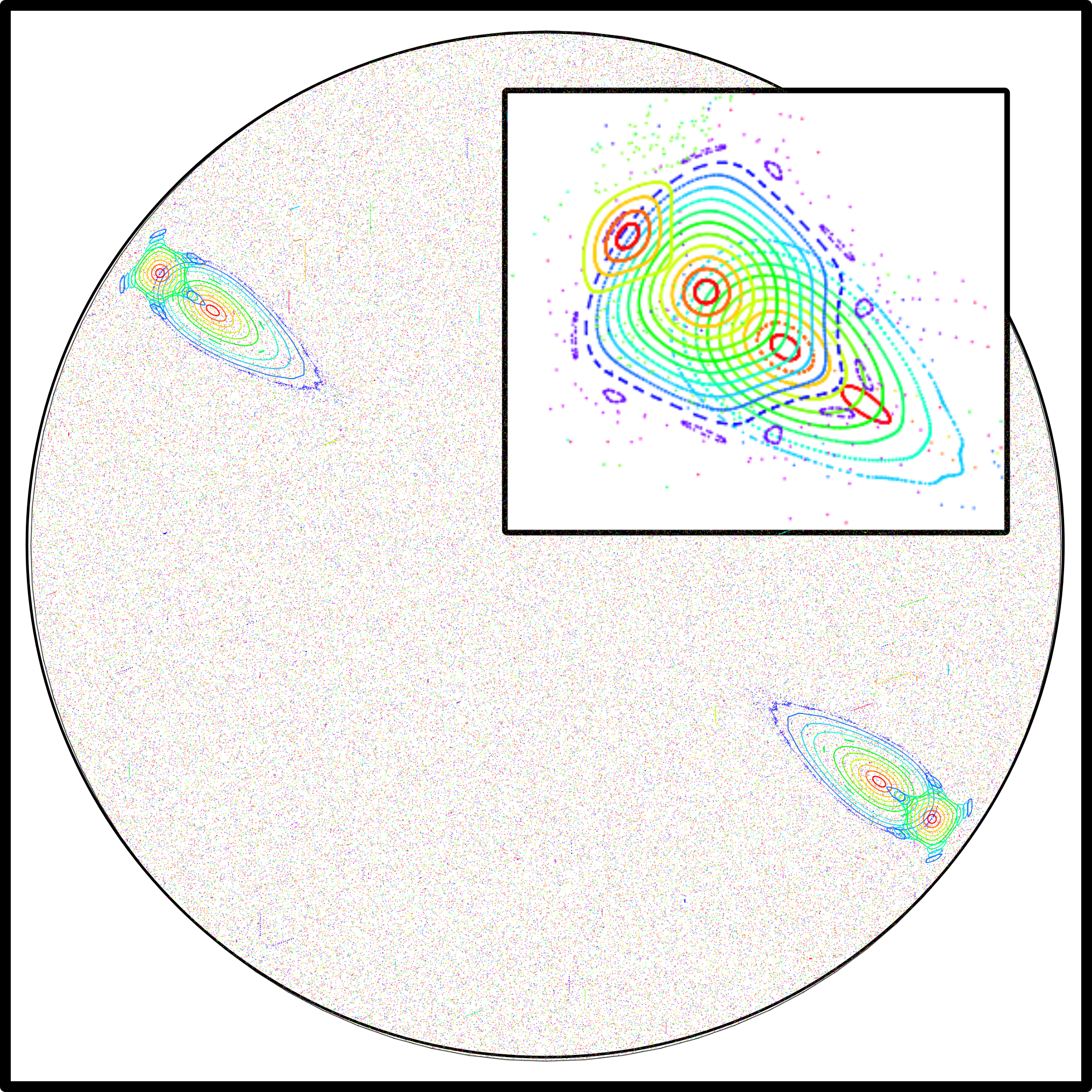}
    \caption{\small Velocity phase projection for Sinai type billiard $S(0.3,\eta_u)$, with the central (low spin, nearly orthogonal) $2$-periodic orbits unstable, while the outer (high spin, nearly tangent) orbits remain stable. For any $\eta>0$ small elliptic islands persist, as shown in the inset, an enlarged view from $S(0.05,\eta_u)$, too small to be visible viewing the full portrait.   }    
\label{fig:sinaizoom}
\end{figure} 

\section{\label{sec:3billiards} Examples}

In this section we consider three illustrative examples generalizing families of standard specular billiards, specifically Sinai dispersers, ellipses, and symmetric three arc billiards which vary from concave dispersers to triangular to three-petal flowers. Each has a table with geometry determined by the first parameter, as introduced in Section \ref{sec:introduction}, and with the mass distribution of the particle corresponding to the second parameter $\eta$. 

\begin{figure}[htbp]
    \centering
    \includegraphics[width=.48\textwidth]{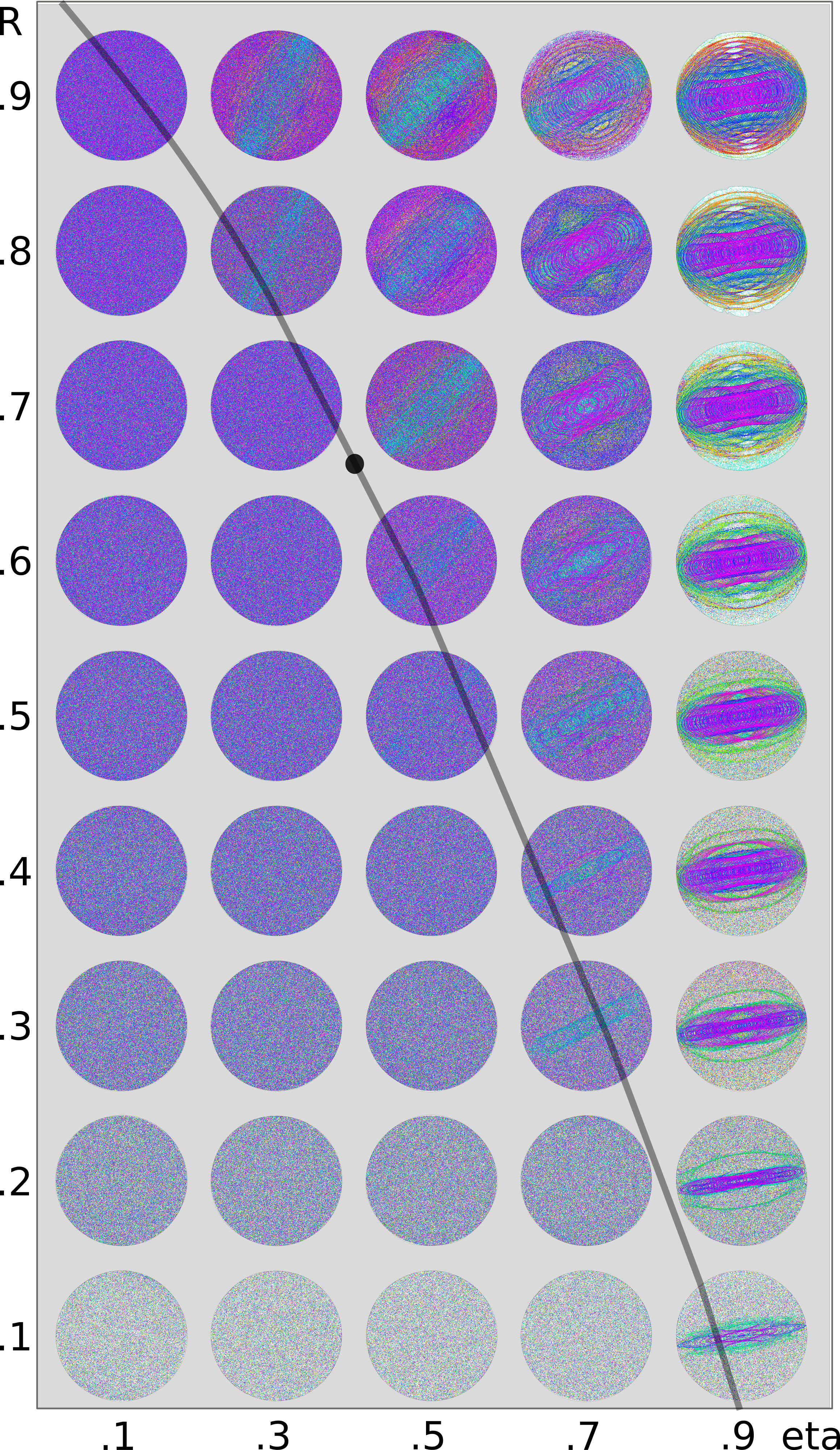}
    \caption{\small Velocity phase projections for Sinai type billiards $S(R,\eta)$ for $R \in \{.1,\dots,.9\}$ and $\eta \in \{.1, .3, .5, .7, .9\}$. 
    The point shows the long known threshold for spinning disks of uniform mass distribution\cite{W} at which the orthogonal $2$-periodic orbit becomes unstable, while the curve shows the more recent generalization.\cite{CFZ} 
    }    
\label{fig:sinaietar}
\end{figure}

The first family includes the one-parameter subfamily $S(R,0)$, corresponding to the well-known Sinai type dispersing billiard. 
However, these are likely the only ergodic billiards in this family, as Equation \ref{eq:stabthresh} implies that linearly stable $2$-periodic orbits persist for nearly tangential collisions with appropriately high spin. 
Only linear, not Lyapunov, stability has been shown analytically, but numerical evidence suggests that invariant regions exist. See Figure \ref{fig:sinaizoom}. 
Looking at velocity phase projections for a sampling of billiards in $S(R,\eta)$ shows an orderly transition from ergodicity at the specular boundary ($\eta=0$) to  completely periodic regularity at the infinite inertia limit ($\eta=1$). See Figure \ref{fig:sinaietar}, in which the stability threshold for the shortest periodic orbits is given as an informal isocline of regularity.     

Recall that a standard planar billiard is said to be (globally) integrable if its phase space is foliated by smooth
closed curves invariant by the billiard map. The ellipse is a well known integrable example, but still much studied as the classical Birkhoff conjecture, which claims that the boundary of a strictly convex integrable billiard table is necessarily an ellipse, remains open.\cite{KS} Our interest is in the extent to which the integrability persists for ellipses in the more general family. The subfamilies $E(0,\eta)$ and $E(e,0)$ are integrable: in the former case, circular no-slip billiards are well understood and have caustics comprised of two circles concentric with the boundary,\cite{CF} while the latter corresponds to standard ellipse billiards. (Strictly speaking, one trivally extends the foliation of invariant curves orthogonally in the rotational direction to obtain a foliation of invariant surfaces for the generalized billiard.)  
\begin{figure}[htbp]
    \centering

    \includegraphics[width=.48\textwidth]{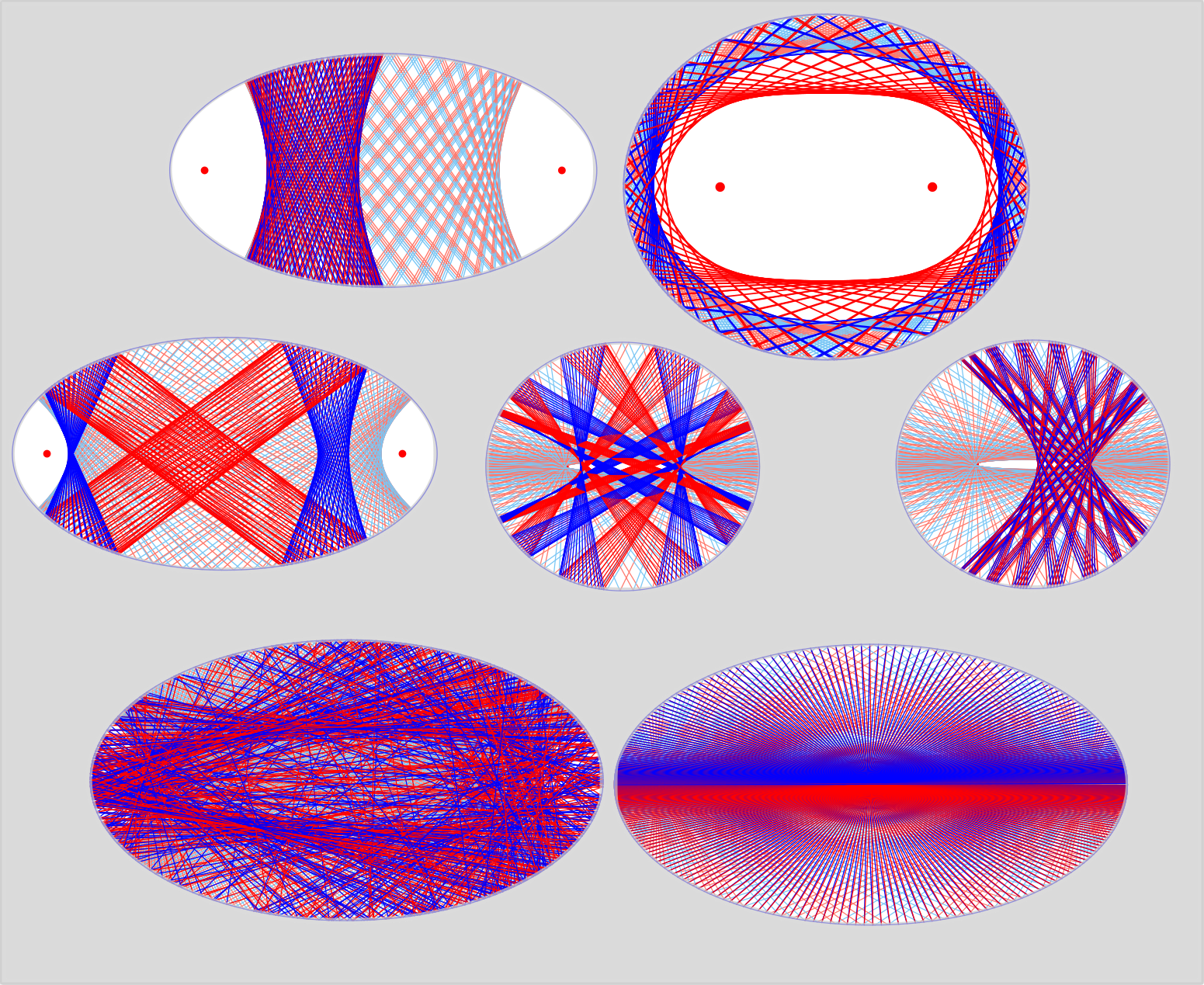}
    \caption{\small Trajectories for the classical ellipse $E(e,0)$ (lighter, background) and $E(e,\eta)$ (darker, foreground) with same initial conditions. Colors alternate between collisions to highlight the structure. Top row: $E(.7,\eta_u)$, $E(.5, \eta_u)$; Middle row: $E(.7,.25)$, $E(.3,.25)$, $E(.3,.5)$; Bottom row: $E(.85,.5)$, $E(.85,.99)$.}    
\label{fig:ellipses}
\end{figure}

In the special case of the circle, when we perturb the classical model given by $\eta=0$ to a collision model with $\eta>0$, the integrability remains with the circular caustics replaces by caustics comprised of pairs of circles (or in the full configuration space, cylinders).
Borisov, Kilin, and Mamaev\cite{nonholonomic} give a numeric example of generic eccentricity which apparently 
contains a chaotic layer suggesting nonintegrability. In general, though, it appears that a variety of dynamic behaviors may be possible. In some instances varying $\eta$ yields a variation which does not destroy the integrability of the classical specular case (for example, splitting the elliptical caustic into two curves or destroying the symmetry of the hyberbolic caustic, as in Figure \ref{fig:ellipses}, top row) but in other cases more complicated dynamics arise, such as elliptic islands around higher order periodic points, Figure \ref{fig:ellipses}, middle row.
It is perhaps worth noting that with the added dimension it is possible that the dynamics may become more complicated topologically without necessarily displaying hyperbolic behavior. 


\begin{figure}[htbp]
    \centering

    \includegraphics[width=.48\textwidth]{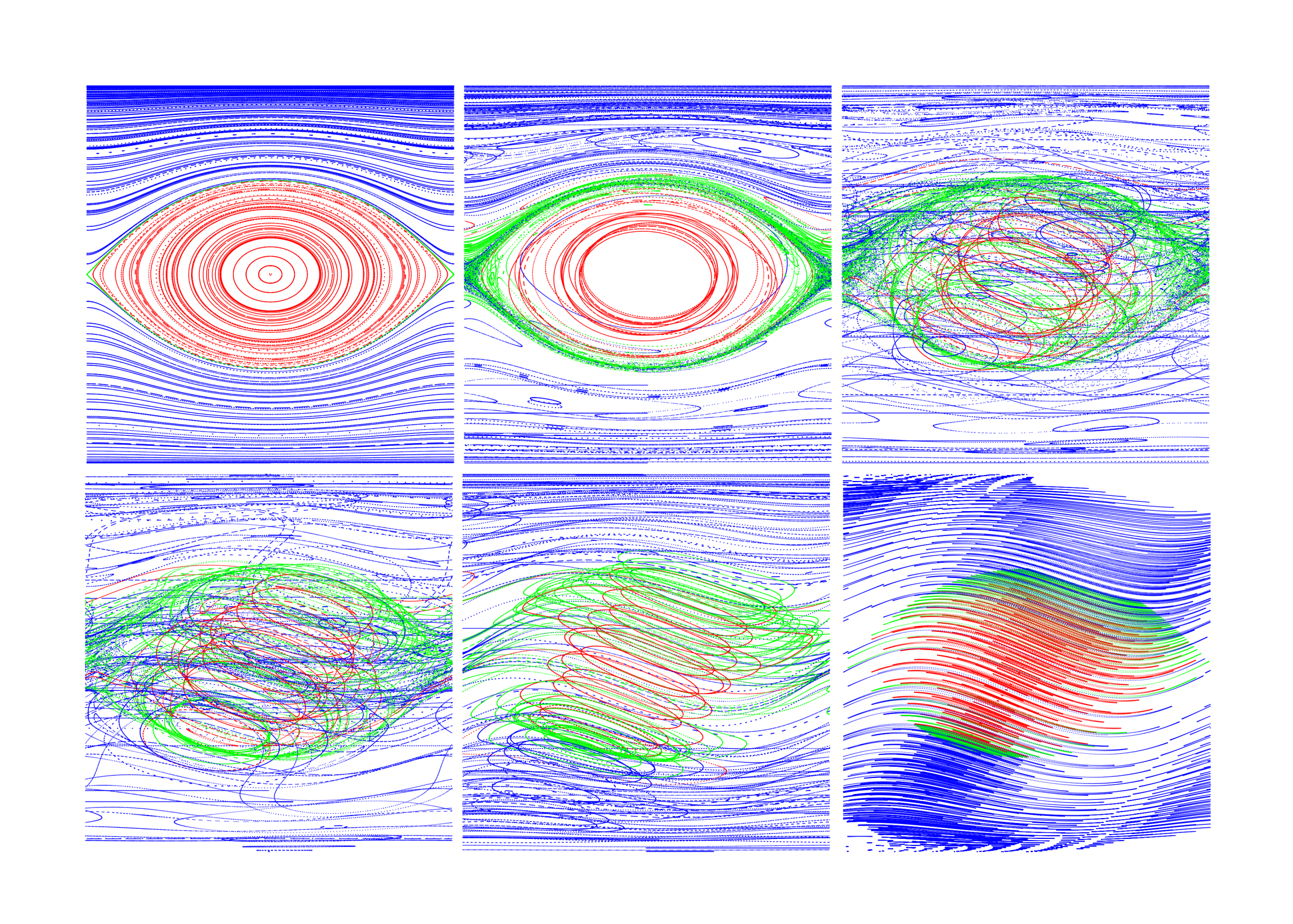}
    \caption{\small Projections to the standard phase space for the ellipse billiard $E(0.5, \eta )$ with $\eta$ varying from $0$ to $0.9999$, showing one of the two identical halves. Varying from the specular billiard to the infinite moment extreme, the integrable structure appears to dissipate in the intermediate cases before a new structure forms.}    
\label{fig:ellipsespec}
\end{figure}

For a more systematic example, we restrict to the subfamily $E(.5,\eta)$ and look at how the phase space evolves as we vary $\eta$, starting with the classical billiards case of $\eta=0$ and ranging to the infinite inertia limit case of $\eta=1$. A view of the progression of the dynamics is given by the projection to standard phase space in Figure \ref{fig:ellipsespec}.


\begin{figure*}[htbp]
    \centering

    \includegraphics[width=.96\textwidth]{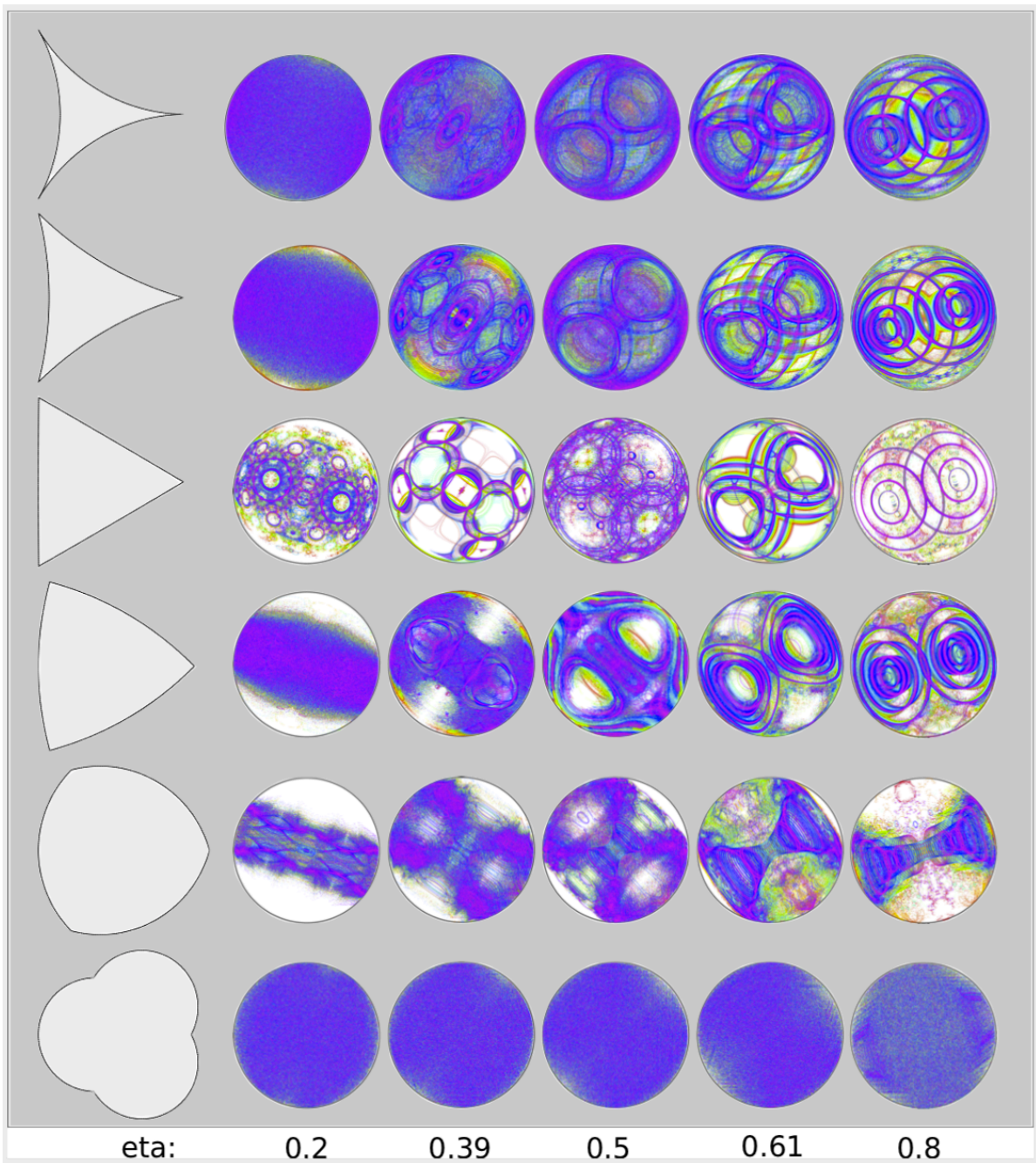}
    \caption{\small In the left column, tables in the family $F(\theta, \eta)$, with velocity phase projections to the right displaying the wide range of possible dynamics. (The values $\eta=.39$ and $\eta=.61$ are chosen for their proximity to the dual persistently periodic $\eta$ values for the triangle.)   }    
\label{fig:3petal}
\end{figure*}

Billiards with boundary comprised of arcs of circles have been widely studied and are known to demonstrate varied dynamics, often including divided phase spaces with elliptic islands scattered throughout an ergodic sea.\cite{lemonergo, moon, BunZhang, umbrellaergodicity}
The subfamily $F(\theta, 0)$ of symmetric standard billiards comprised of equal curvature arcs connecting equilateral vertices is a microcosm of this variety, and the generalized family is accordingly diverse as $\eta$ increases (Figure \ref{fig:3petal}). For negative values of $\theta$, the dispersing billiards in the specular limit quickly become mixed as the particle inertia increases and stable periodic points emerge. In contrast, for the larger values of $\theta$ the defocusing mechanism appears to maintain the ergodicity created in the standard case for all but the largest values of $\eta$. Unlike for the Sinai family $S(R,\eta)$, we know of no small scale invariant regions which would preclude ergodicity but would not be visible on the standard scale.  This is of interest, as currently no no-slip billiard has been analytically shown to be ergodic, and lacking linearly stable $2$-periodic orbits these billiards appear to be candidates.

\section{\label{sec:pp}Persistent Periodicity}
The study of periodic orbits often plays a central role in understanding billiards, whether one is interested physical applications, understanding periodicity for its own sake, or even destroying elliptic periodic points to create ergodic billiards.
It was in the first context (pondering the question of whether one can ``hear the shape of a drum''\cite{Kac}) that  Ivrii conjectured that for every billiard in Euclidean space the set of periodic orbits has measure zero.\cite{ivrii} The conjecture remains open and accordingly no $k$-reflective billiards, that is,  billiards with phase space having an open set comprised entirely of $k$-periodic points, have been found for Euclidean domains.\cite{fierobe2021thesis} 

Of interest to us is the even stronger condition, persistent periodicity, in which the periodic orbits comprise a full measure subset of phase space. While it appears that even the weaker $k$-reflective condition may not exist in the standard case, some examples of persistent periodicity have been found in variant billiards. In addition to the no-slip triangle,\cite{CFZ} examples have been found on non-Euclidean domains including the $2$-sphere\cite{reflectivesphere} and for alternative collision models including projective\cite{fierobe2020projective} and symplectic billiards.\cite{albers2019polygonal} Theorem \ref{thm:main}, which we now prove, shows that in at least two senses dense families of persistently periodic no-slip billiards exist.

\subsection{\label{ssec:stripwedge}Proof of Theorem \ref{thm:main}}

The fact that the no-slip strip is bounded is an elementary though significant result, and for the uniform case has been proved through a direct summation\cite{gutkin} as well as by a more geometric proof observing that the projection into the tangential rotational plane is the envelope of an astroid.\cite{CFII} Both arguments easily generalize to give the following lemma, which we use below.
\begin{lemma}
For the no-slip strip with particles corresponding to any $\eta$, any (non-horizontal) trajectory is bounded.
\label{lm:bounded}
\end{lemma}

\noindent
First we show that for any $n$ there is a choice of $\eta$ for which all trajectories of the no-slip strip will be $2n$-periodic.
Let $T$ be the transformation matrix given in Equation \ref{eq:T}.
Here and subsequently we define a rotational matrix  \[R_{\theta} =
\begin{pmatrix} 1 & 0 & 0
\\ 0 & \cos\theta & -\sin\theta
\\ 0 & \sin\theta & \cos\theta
\\ \end{pmatrix},\]
which for the change of basis in the strip will merely be \[R_{\pi} =
\begin{pmatrix} 1 & 0 & 0
\\ 0 & - 1 & 0
\\ 0 & 0 & - 1
\\ \end{pmatrix}.\] Then for starting velocity $\textbf{v}^-$in the fixed frame $(e_1,e_2,e_3)$ and $\textbf{v}^+$ the return vector after two collisions, $$\textbf{v}^+ = \Phi^{2}(\textbf{v}^-) $$
where 
\[\Phi^{2} = TR_{\pi}TR_{\pi}= \begin{pmatrix} \cos\left( 2\pi\eta \right) & - \sin\left( 2\pi\eta \right) & 0
\\ \sin\left( 2\pi\eta \right) & \cos\left( 2\pi\eta \right) & 0
\\ 0 & 0 & 1
\\ \end{pmatrix}.\] 
\begin{figure}[htbp]
    \centering

    \includegraphics[width=.48\textwidth]{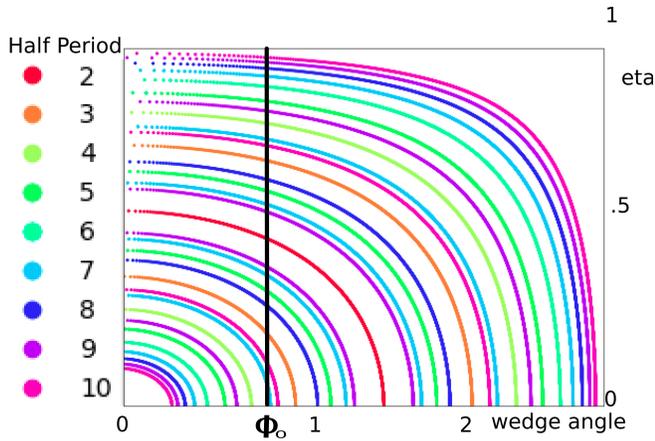}
    \caption{\small Curves indicating combinations of wedge angle $2\phi$ and mass distribution $\eta$ which yield persistently periodic no-slip wedges of rotational period $2$ through $10$, yielding trajectories of periods $4$ through $20$. For any wedge angle smaller than $2\phi_0$, a value of $\eta$ exists yielding $2n$ persistent periodicity for any $n \geq 2$.
    }
    \label{fig:wedgeper}
\end{figure}
By induction,  for any positive integer \(n\)
\begin{equation}
\label{eqn:phiToN}
\Phi^{2n} = \begin{pmatrix}
\cos\left( 2\pi n\eta \right) & - \sin\left( 2\pi n\eta \right) & 0 \\
\sin\left( 2\pi n\eta \right) & \cos\left( 2\pi n\eta \right) & 0 \\
0 & 0 & 1 \\
\end{pmatrix}
\end{equation}
and \(\Phi^{2n} = I\) if and only if
\(\cos\left( 2\pi n\eta \right) = 1\), in which case 
\(\sin\left( 2\pi n\eta \right) = 0\). For any positive integer
\(n\), if \(\eta = \frac{m}{n}\) the velocity map on the no-slip wedge will be periodic  for any integer \(0<m<n\).
If there were a drift in position after $n$ iterations of the return map, it would be repeated for each subsequent $n$ iterations and the trajectories would be unbounded, violating Lemma \ref{lm:bounded}, hence the position is also periodic. 
This proves (1), as well as the initial assertion about the distribution of potentially persistently periodic $\eta$.

Even with only two sides, the changing frames makes the straightforward approach we use for the strip computationally daunting for the wedge. However, if we consider rotational coordinates Equation \ref{eq:rotate} yields an implicit relation between the wedge angle $2\phi$ and $\eta$ for any fixed rational rotational angle $\theta=\frac{m}{n}$, as  in Figure \ref{fig:wedgeper}. (Recall that $\theta$ represents the rotation of the velocity vector about the eigenvector of the transformation matrix in the tangent space, not the physical rotation of he particle.) For the proof it is sufficient to look at a single branch of the solutions given by the quartic Equation \ref{eq:rotate}. Specifically, for a given $\phi$, an $\eta$ yielding rotational period $\theta=\frac{m}{n}\pi$ will exist if it satisfies
\begin{equation}
    \cos^2\left( \frac{\pi \eta }{2}\right)=
    \frac{\sqrt{2}+\sqrt{1+\cos \left(\frac{m}{n}\pi\right) }}{2\sqrt{2}\cos^2\phi},
    \label{eq:onebranch}
\end{equation}
which will hold as long as 
\begin{equation}
    \sqrt{2}+\sqrt{1+\cos\left(\frac{m}{n}\pi\right)} \leq 2\sqrt{2} \cos^2\phi.
    \label{eq:rotlim}
\end{equation}
We need only shown that for any $n \geq 2$ there exists some $m$ yielding a solution, and since for any $n>2$ we may choose $m=n-1>1$ so that $\cos \theta <0$, if Equation \ref{eq:rotlim} is satisifed for $m=1$ and $n=2$ for a given $\phi$ we can find a solution for any $n$. Letting $\frac{m}{n}=\frac{1}{2}$ we see that it holds for any $$\phi \leq \arccos{\frac{\sqrt{2+\sqrt{2}}}{2}}=\frac{\pi}{8}, $$ and it follows that we may choose $\eta$ to obtain periodic velocity as 
long as we have wedge angle $2\phi \leq \frac{\pi}{4}$. 
Recent boundeness results for the wedge ensure that the position must also be periodic, using the same drifting prohibition as was employed in the case of the strip.
This proves (2), completing the proof of Theorem \ref{thm:main}.

\begin{figure}[htbp]
    \centering

    \includegraphics[width=.48\textwidth]{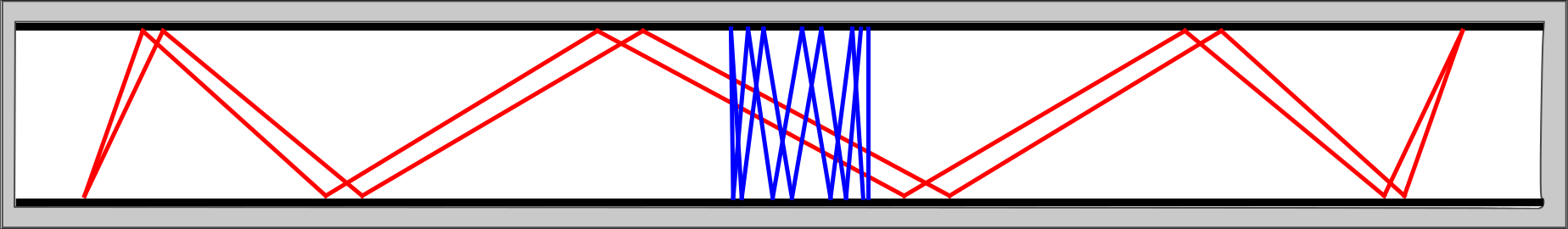}
    \caption{\small For the infinite strip, the mass distribution of the particle may be chosen so that all trajectories are $2n$-periodic for any positive integer $n>1$. Here $n=7$.
    }
    \label{fig:strip}
\end{figure}

Simulations confirm the results of the theorem, as for example the persistently $14$-periodic  strip shown  in Figure \ref{fig:strip}. Numerical results suggest that persistently period wedges are common for angles greater than $\frac{\pi}{4}$, however our searches have also failed for some large angles, and the question of the distribution of persistently periodic wedges of larger angles remains open.

\subsection{\label{ssec:ppp}Persistently periodic polygons}

The equilateral triangle no-slip billiard with particles of uniform mass distribution ($\eta_u \approx 0.392$) is known to be persistently periodic,\cite{CFZ} with the phase space comprised of $4$-reflective and $6$-reflective sets of combined full measure. 
We present new examples here, found in a numerical search of mass distributions for equilateral polygons.

For a regular $n$-gon, consider the set of numbered sides $S=\{0,1,\dots, n-1\}$. For any $k \geq 3$ let $\mathcal{I} $ be a length $k$ sequence of integers in $S$, allowing nonconsecutive repetitions. 
Then the velocity transformation of the no-slip billiard map for a given $\eta$ of an orbit colliding with side corresponding to $I$ is given by
$$ \Phi_{\eta}^{\mathcal{I} }= \prod_\mathcal{I}  R_{\theta_i}TR_{-\theta_i},$$
where $\theta_i$ is the angle between the sides given by entries $i$ and $i-1$ in $\mathcal{I}$ and the transformation and rotational matrices are as defined above. 
Define an error function $$E_\mathcal{I} (\eta)=\| \Phi_{\eta}^{\mathcal{I} }-I \|$$ where $I$ is the identity matrix and the norm is the Euclidean norm viewing the matrices in $\mathbb{R}^9$. 
Then for a billiard to be $k$-reflective a necessary condition is that $E_\mathcal{I}(\eta)=0$ for some sequence $\mathcal{I}$ of length $k$. 
Note that precluding repetition of a side is not sufficient to ensure that a sequence is dynamically allowable. 
Additionally, even among dynamically allowable length $k$ sequences, a nonzero error does not preclude persistent periodicity, since the sequence may correspond to a $l$-reflective component of phase space for some $l<k$. 
Nonetheless, candidates will necessarily produce zero error for some sequences, and it is straightforward procedure for a given polynomial to check a sampling of $\eta$ values cycling through the combinatorial possibilities for relatively small $k$ values.

\begin{figure}[htbp]
    \centering

    \includegraphics[width=.45\textwidth]{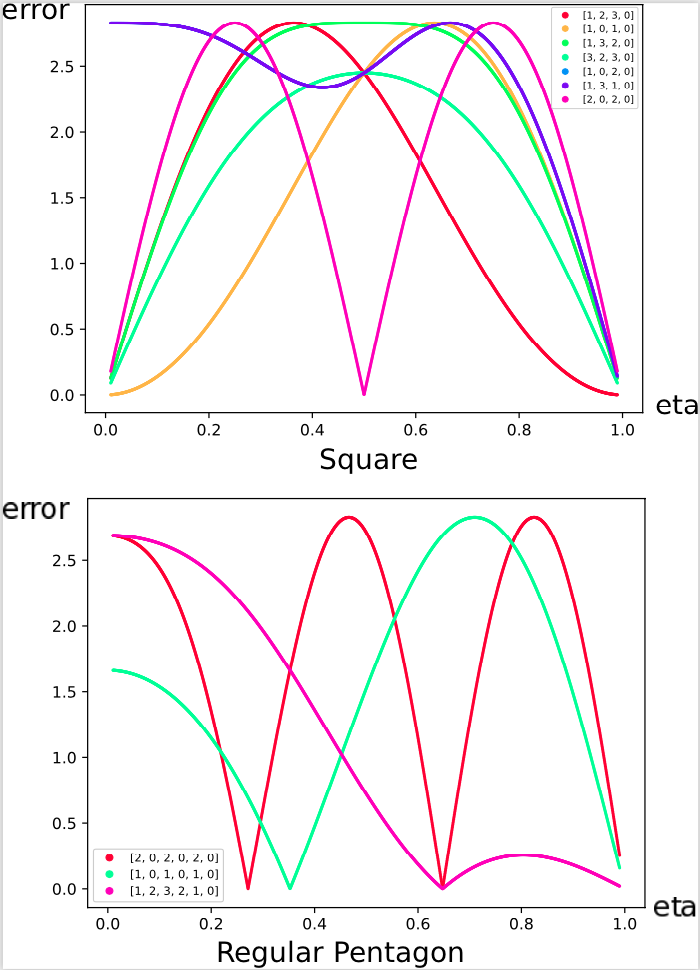}
    \caption{\small Numerically checking the difference between the billiard map composition and the identity, for the given wall sequences, yields candidate values of $\eta$ to result in persistant periodicity. For the square, $\eta=0.5$ can be verified, while two of the three candidates for the regular pentagon can be verified.   }    
\label{fig:ppsearch}
\end{figure}

Note that $E_{\mathcal{I}}(1)=0$ for any $\mathcal{I}$, since all billiards are are persistently periodic trivially for $\eta=1$, with all orbits $2$-periodic. 
Examples of the search results for the square for some sequences with $k=4$ and the pentagon with $k=6$ are given in Figure \ref{fig:ppsearch}. In the square case, $\eta=0.5$ emerges as a candidate, while three candidates emerge for the pentagon. It is straightforward to numerically check whether the candidate $\eta$ values result in persistently periodic billiards, though an analytic proof requires a case analysis which we do not give here.\cite{CF} Two of the candidates for the pentagon yield persistent periodicity,
as well as $\eta=0.5$ for the square as shown in Figure \ref{fig:PPSquare}. The search also turned up a second $\eta$ value for which the equilateral triangle is persistently period, which combined with the pentagon $\eta$ values suggests a duality with symmetric distributions inside and beyond the radius of collision. The results are summarized in Table \ref{tab:PPP}. Polygons with six or more sides were also considered, but no persistently periodic examples were found, indicating that the periods were higher than the scope of our search or possibly no such examples exist.

\begin{table}
\label{tab:PPP}
\centering
\begin{tabular}{||c||c|c||} 
\hhline{~|t:==|}
\multicolumn{1}{c||}{} & $\eta$ & possible periods\footnote{Not including degeneracies from overlapping trajectories  }  \\ 
\hhline{|t:=::==|}
Equilateral            &  $\frac{1}{\pi}\arccos{\frac{1}{3}} 
\approx 0.3918$  &        4, 6            \\ 
\cline{2-3}
Triangle               &  $1-\frac{1}{\pi}\arccos{\frac{1}{3}} \approx 0.6082$    &          6, 12         \\ 
\hline
Square                 & 0.5 &     4, 6, 8 10, 12              \\ 
\hline
Regular                & $\frac{2}{\pi}\arccos \left( \frac{\sqrt{1+\cos\frac{2\pi}{3}}}{\sqrt{2}\cos \frac{3\pi}{10}}\right)\approx$ 0.35     &       \makecell{6, 8, 10, 14, \\ 20, 30, 40}            \\ 
\cline{2-3}
Pentagon               & $1-\frac{2}{\pi}\arccos \left( \frac{\sqrt{1+\cos\frac{2\pi}{3}}}{\sqrt{2}\cos \frac{3\pi}{10}}\right) \approx$ 0.65     &        \makecell{6, 10, 20, \\ 30, 40}           \\
\hhline{|b:=:b:==|}
\end{tabular}
\end{table}

\section{Concluding Remarks}

Many of the natural applications and problems motivating the study of billiards have natural analogues allowing for rotating particles with positive inertia, 
suggesting that this generalized framework may prove useful. 
Even if one's focus is on standard specular billiards, it may be possible to glean new insights by viewing them as a limit in this setting.  The question of which dynamical properties found for small $\eta$ persist in the limit case, however, is unclear. For example, the ubiquitous periodic points for $\eta>0$ no-slip billiards will dissipate unless the rotational velocity becomes infinitely large relative to the linear velocities.

Finally, it's worth noting that even with the flexibility we allow in the mass distribution of the particles, our model is still highly constrained. The underlying rigid body collision theory can be developed very generally for particles described by a broad class of measurable sets,\cite{CF} and it seems likely that many options remain for relatively simple billiard systems which may demonstrate interesting dynamics. For example, one might retain the requirement of disk-shaped colliding particles but eliminate the radial symmetry, an idea that has yielded interesting alternatives in the design of bowling balls. 

\section{Acknowledgements}
We would like to thank the University of Delaware for their support of
Jan Ahmed and Bill Wang through the Summer Scholars program.  We would like to thank Frank Morgan for his useful comments and questions.

The persistently periodic square for $\eta=\frac{1}{2}$ was discovered independently by Clayton Boone and Bishwas Ghimire in a mathematical modeling class at Tarleton State University. Ghimire also contributed several useful ideas used here to further study persistently periodic polygons.

\bibliography{VaryingMass}

\begin{thebibliography}{35}%
\makeatletter
\providecommand \@ifxundefined [1]{%
 \@ifx{#1\undefined}
}%
\providecommand \@ifnum [1]{%
 \ifnum #1\expandafter \@firstoftwo
 \else \expandafter \@secondoftwo
 \fi
}%
\providecommand \@ifx [1]{%
 \ifx #1\expandafter \@firstoftwo
 \else \expandafter \@secondoftwo
 \fi
}%
\providecommand \natexlab [1]{#1}%
\providecommand \enquote  [1]{``#1''}%
\providecommand \bibnamefont  [1]{#1}%
\providecommand \bibfnamefont [1]{#1}%
\providecommand \citenamefont [1]{#1}%
\providecommand \href@noop [0]{\@secondoftwo}%
\providecommand \href [0]{\begingroup \@sanitize@url \@href}%
\providecommand \@href[1]{\@@startlink{#1}\@@href}%
\providecommand \@@href[1]{\endgroup#1\@@endlink}%
\providecommand \@sanitize@url [0]{\catcode `\\12\catcode `\$12\catcode
  `\&12\catcode `\#12\catcode `\^12\catcode `\_12\catcode `\%12\relax}%
\providecommand \@@startlink[1]{}%
\providecommand \@@endlink[0]{}%
\providecommand \url  [0]{\begingroup\@sanitize@url \@url }%
\providecommand \@url [1]{\endgroup\@href {#1}{\urlprefix }}%
\providecommand \urlprefix  [0]{URL }%
\providecommand \Eprint [0]{\href }%
\providecommand \doibase [0]{http://dx.doi.org/}%
\providecommand \selectlanguage [0]{\@gobble}%
\providecommand \bibinfo  [0]{\@secondoftwo}%
\providecommand \bibfield  [0]{\@secondoftwo}%
\providecommand \translation [1]{[#1]}%
\providecommand \BibitemOpen [0]{}%
\providecommand \bibitemStop [0]{}%
\providecommand \bibitemNoStop [0]{.\EOS\space}%
\providecommand \EOS [0]{\spacefactor3000\relax}%
\providecommand \BibitemShut  [1]{\csname bibitem#1\endcsname}%
\let\auto@bib@innerbib\@empty
\bibitem [{\citenamefont {Chumley}\ \emph {et~al.}(2021)\citenamefont
  {Chumley}, \citenamefont {Feres},\ and\ \citenamefont
  {Garcia~German}}]{chumley}%
  \BibitemOpen
  \bibfield  {author} {\bibinfo {author} {\bibfnamefont {T.}~\bibnamefont
  {Chumley}}, \bibinfo {author} {\bibfnamefont {R.}~\bibnamefont {Feres}}, \
  and\ \bibinfo {author} {\bibfnamefont {L.~A.}\ \bibnamefont
  {Garcia~German}},\ }\href {\doibase 10.1137/20M1349552} {\bibfield  {journal}
  {\bibinfo  {journal} {SIAM Journal on Applied Dynamical Systems}\ }\textbf
  {\bibinfo {volume} {20}},\ \bibinfo {pages} {1655} (\bibinfo {year}
  {2021})},\ \Eprint {http://arxiv.org/abs/https://doi.org/10.1137/20M1349552}
  {https://doi.org/10.1137/20M1349552} \BibitemShut {NoStop}%
\bibitem [{\citenamefont {Baryakhtar}\ \emph {et~al.}(2006)\citenamefont
  {Baryakhtar}, \citenamefont {Yanovsky}, \citenamefont {Naydenov},\ and\
  \citenamefont {Kurilo}}]{composite}%
  \BibitemOpen
  \bibfield  {author} {\bibinfo {author} {\bibfnamefont {V.~G.}\ \bibnamefont
  {Baryakhtar}}, \bibinfo {author} {\bibfnamefont {V.~V.}\ \bibnamefont
  {Yanovsky}}, \bibinfo {author} {\bibfnamefont {S.~V.}\ \bibnamefont
  {Naydenov}}, \ and\ \bibinfo {author} {\bibfnamefont {A.~V.}\ \bibnamefont
  {Kurilo}},\ }\href {\doibase 10.1134/S1063776106080127} {\bibfield  {journal}
  {\bibinfo  {journal} {Journal of Experimental and Theoretical Physics}\
  }\textbf {\bibinfo {volume} {103}},\ \bibinfo {pages} {292} (\bibinfo {year}
  {2006})}\BibitemShut {NoStop}%
\bibitem [{\citenamefont {Datseris}\ \emph {et~al.}(2019)\citenamefont
  {Datseris}, \citenamefont {Hupe},\ and\ \citenamefont
  {Fleischmann}}]{Datseris}%
  \BibitemOpen
  \bibfield  {author} {\bibinfo {author} {\bibfnamefont {G.}~\bibnamefont
  {Datseris}}, \bibinfo {author} {\bibfnamefont {L.}~\bibnamefont {Hupe}}, \
  and\ \bibinfo {author} {\bibfnamefont {R.}~\bibnamefont {Fleischmann}},\
  }\href {\doibase 10.1063/1.5099446} {\bibfield  {journal} {\bibinfo
  {journal} {Chaos: An Interdisciplinary Journal of Nonlinear Science}\
  }\textbf {\bibinfo {volume} {29}},\ \bibinfo {pages} {093115} (\bibinfo
  {year} {2019})},\ \Eprint
  {http://arxiv.org/abs/https://doi.org/10.1063/1.5099446}
  {https://doi.org/10.1063/1.5099446} \BibitemShut {NoStop}%
\bibitem [{\citenamefont {Davis}\ \emph {et~al.}(2018)\citenamefont {Davis},
  \citenamefont {DiPietro}, \citenamefont {Rustad},\ and\ \citenamefont
  {Laurent}}]{Davis}%
  \BibitemOpen
  \bibfield  {author} {\bibinfo {author} {\bibfnamefont {D.}~\bibnamefont
  {Davis}}, \bibinfo {author} {\bibfnamefont {K.}~\bibnamefont {DiPietro}},
  \bibinfo {author} {\bibfnamefont {J.}~\bibnamefont {Rustad}}, \ and\ \bibinfo
  {author} {\bibfnamefont {A.~S.}\ \bibnamefont {Laurent}},\ }\href {\doibase
  doi:10.1515/advgeom-2017-0053} {\bibfield  {journal} {\bibinfo  {journal}
  {Advances in Geometry}\ }\textbf {\bibinfo {volume} {18}},\ \bibinfo {pages}
  {133} (\bibinfo {year} {2018})}\BibitemShut {NoStop}%
\bibitem [{\citenamefont {Blasi}\ and\ \citenamefont
  {Terracini}(2021)}]{deblasi}%
  \BibitemOpen
  \bibfield  {author} {\bibinfo {author} {\bibfnamefont {I.~D.}\ \bibnamefont
  {Blasi}}\ and\ \bibinfo {author} {\bibfnamefont {S.}~\bibnamefont
  {Terracini}},\ }\href@noop {} {\enquote {\bibinfo {title} {Refraction
  periodic trajectories in central mass galaxies},}\ } (\bibinfo {year}
  {2021}),\ \Eprint {http://arxiv.org/abs/2105.02108} {arXiv:2105.02108
  [math.DS]} \BibitemShut {NoStop}%
\bibitem [{\citenamefont {Bunimovich}(2019)}]{Bun2}%
  \BibitemOpen
  \bibfield  {author} {\bibinfo {author} {\bibfnamefont {L.~A.}\ \bibnamefont
  {Bunimovich}},\ }\href {\doibase 10.1063/1.5122195} {\bibfield  {journal}
  {\bibinfo  {journal} {Chaos}\ }\textbf {\bibinfo {volume} {29}},\ \bibinfo
  {pages} {091105, 6} (\bibinfo {year} {2019})}\BibitemShut {NoStop}%
\bibitem [{\citenamefont {Cox}\ and\ \citenamefont {Feres}(2016)}]{CF}%
  \BibitemOpen
  \bibfield  {author} {\bibinfo {author} {\bibfnamefont {C.}~\bibnamefont
  {Cox}}\ and\ \bibinfo {author} {\bibfnamefont {R.}~\bibnamefont {Feres}},\
  }\href {\doibase 10.3934/dcds.2016065} {\bibfield  {journal} {\bibinfo
  {journal} {Discrete Contin. Dyn. Syst.}\ }\textbf {\bibinfo {volume} {36}},\
  \bibinfo {pages} {6065} (\bibinfo {year} {2016})}\BibitemShut {NoStop}%
\bibitem [{Note1()}]{Note1}%
  \BibitemOpen
  \bibinfo {note} {The model arising as the small radius limit of the no-slip
  (or ``rough'') alternative gives a complete reversal of rotational velocity
  at collisions, while the specular case is generally thought to be
  frictionless and accordingly leaving rotation unchanged. Projecting out the
  rotation, however, yields identical models.}\BibitemShut {Stop}%
\bibitem [{\citenamefont {Cowan}(2008)}]{cowan}%
  \BibitemOpen
  \bibfield  {author} {\bibinfo {author} {\bibfnamefont {D.}~\bibnamefont
  {Cowan}},\ }\href@noop {} {\bibfield  {journal} {\bibinfo  {journal}
  {Discrete and Continuous Dynamical Systems}\ }\textbf {\bibinfo {volume}
  {22}},\ \bibinfo {pages} {101} (\bibinfo {year} {2008})}\BibitemShut
  {NoStop}%
\bibitem [{\citenamefont {Baryshnikov}\ \emph {et~al.}(2014)\citenamefont
  {Baryshnikov}, \citenamefont {Blumen}, \citenamefont {Kim},\ and\
  \citenamefont {Zharnitsky}}]{baryshnikov}%
  \BibitemOpen
  \bibfield  {author} {\bibinfo {author} {\bibfnamefont {Y.}~\bibnamefont
  {Baryshnikov}}, \bibinfo {author} {\bibfnamefont {V.}~\bibnamefont {Blumen}},
  \bibinfo {author} {\bibfnamefont {K.}~\bibnamefont {Kim}}, \ and\ \bibinfo
  {author} {\bibfnamefont {V.}~\bibnamefont {Zharnitsky}},\ }\href {\doibase
  https://doi.org/10.1016/j.physd.2013.11.007} {\bibfield  {journal} {\bibinfo
  {journal} {Physica D: Nonlinear Phenomena}\ }\textbf {\bibinfo {volume}
  {269}},\ \bibinfo {pages} {21} (\bibinfo {year} {2014})}\BibitemShut
  {NoStop}%
\bibitem [{Note2()}]{Note2}%
  \BibitemOpen
  \bibinfo {note} {Since the no-slip model depends only on the moment of
  inertia, each $\eta $ actually corresponds to an equivalence class of mass
  distributions.}\BibitemShut {Stop}%
\bibitem [{\citenamefont {Cox}\ and\ \citenamefont {Feres}(2017)}]{CFII}%
  \BibitemOpen
  \bibfield  {author} {\bibinfo {author} {\bibfnamefont {C.}~\bibnamefont
  {Cox}}\ and\ \bibinfo {author} {\bibfnamefont {R.}~\bibnamefont {Feres}},\
  }in\ \href {\doibase 10.1090/conm/698/14032} {\emph {\bibinfo {booktitle}
  {Dynamical systems, ergodic theory, and probability: in memory of {K}olya
  {C}hernov}}},\ \bibinfo {series} {Contemp. Math.}, Vol.\ \bibinfo {volume}
  {698}\ (\bibinfo  {publisher} {Amer. Math. Soc., Providence, RI},\ \bibinfo
  {year} {2017})\ pp.\ \bibinfo {pages} {91--110}\BibitemShut {NoStop}%
\bibitem [{\citenamefont {Garwin}(1969)}]{garwin}%
  \BibitemOpen
  \bibfield  {author} {\bibinfo {author} {\bibfnamefont {R.~L.}\ \bibnamefont
  {Garwin}},\ }\href {\doibase 10.1119/1.1975420} {\bibfield  {journal}
  {\bibinfo  {journal} {American Journal of Physics}\ }\textbf {\bibinfo
  {volume} {37}},\ \bibinfo {pages} {88} (\bibinfo {year} {1969})}\BibitemShut
  {NoStop}%
\bibitem [{\citenamefont {Broomhead}\ and\ \citenamefont
  {Gutkin}(1993)}]{gutkin}%
  \BibitemOpen
  \bibfield  {author} {\bibinfo {author} {\bibfnamefont {D.~S.}\ \bibnamefont
  {Broomhead}}\ and\ \bibinfo {author} {\bibfnamefont {E.}~\bibnamefont
  {Gutkin}},\ }\href {\doibase 10.1016/0167-2789(93)90205-F} {\bibfield
  {journal} {\bibinfo  {journal} {Phys. D}\ }\textbf {\bibinfo {volume} {67}},\
  \bibinfo {pages} {188} (\bibinfo {year} {1993})}\BibitemShut {NoStop}%
\bibitem [{\citenamefont {Mej{\'i}a-Monasterio}\ \emph
  {et~al.}(2001)\citenamefont {Mej{\'i}a-Monasterio}, \citenamefont
  {Larralde},\ and\ \citenamefont {Leyvraz}}]{MLL}%
  \BibitemOpen
  \bibfield  {author} {\bibinfo {author} {\bibfnamefont {C.}~\bibnamefont
  {Mej{\'i}a-Monasterio}}, \bibinfo {author} {\bibfnamefont {H.}~\bibnamefont
  {Larralde}}, \ and\ \bibinfo {author} {\bibfnamefont {F.}~\bibnamefont
  {Leyvraz}},\ }\href@noop {} {\bibfield  {journal} {\bibinfo  {journal}
  {Physical review letters}\ }\textbf {\bibinfo {volume} {86 24}},\ \bibinfo
  {pages} {5417} (\bibinfo {year} {2001})}\BibitemShut {NoStop}%
\bibitem [{\citenamefont {Arnol\cprime~d}(1963)}]{Arnold}%
  \BibitemOpen
  \bibfield  {author} {\bibinfo {author} {\bibfnamefont {V.~I.}\ \bibnamefont
  {Arnol\cprime~d}},\ }\href@noop {} {\bibfield  {journal} {\bibinfo  {journal}
  {Uspehi Mat. Nauk}\ }\textbf {\bibinfo {volume} {18}},\ \bibinfo {pages} {13}
  (\bibinfo {year} {1963})}\BibitemShut {NoStop}%
\bibitem [{\citenamefont {Borisov}\ \emph
  {et~al.}(2011{\natexlab{a}})\citenamefont {Borisov}, \citenamefont {Kilin},\
  and\ \citenamefont {Mamaev}}]{BKM}%
  \BibitemOpen
  \bibfield  {author} {\bibinfo {author} {\bibfnamefont {A.~V.}\ \bibnamefont
  {Borisov}}, \bibinfo {author} {\bibfnamefont {A.~A.}\ \bibnamefont {Kilin}},
  \ and\ \bibinfo {author} {\bibfnamefont {I.~S.}\ \bibnamefont {Mamaev}},\
  }\href {\doibase 10.1134/S1560354711060062} {\bibfield  {journal} {\bibinfo
  {journal} {Regul. Chaotic Dyn.}\ }\textbf {\bibinfo {volume} {16}},\ \bibinfo
  {pages} {653} (\bibinfo {year} {2011}{\natexlab{a}})}\BibitemShut {NoStop}%
\bibitem [{\citenamefont {Cox}\ \emph {et~al.}(2021)\citenamefont {Cox},
  \citenamefont {Feres},\ and\ \citenamefont {Zhao}}]{rolling}%
  \BibitemOpen
  \bibfield  {author} {\bibinfo {author} {\bibfnamefont {C.}~\bibnamefont
  {Cox}}, \bibinfo {author} {\bibfnamefont {R.}~\bibnamefont {Feres}}, \ and\
  \bibinfo {author} {\bibfnamefont {B.}~\bibnamefont {Zhao}},\ }\href {\doibase
  10.1134/S1560354721010019} {\bibfield  {journal} {\bibinfo  {journal} {Regul.
  Chaotic Dyn.}\ }\textbf {\bibinfo {volume} {26}},\ \bibinfo {pages} {1}
  (\bibinfo {year} {2021})}\BibitemShut {NoStop}%
\bibitem [{\citenamefont {Chumley}\ \emph {et~al.}(2020)\citenamefont
  {Chumley}, \citenamefont {Cook}, \citenamefont {Cox},\ and\ \citenamefont
  {Feres}}]{CCCF}%
  \BibitemOpen
  \bibfield  {author} {\bibinfo {author} {\bibfnamefont {T.}~\bibnamefont
  {Chumley}}, \bibinfo {author} {\bibfnamefont {S.}~\bibnamefont {Cook}},
  \bibinfo {author} {\bibfnamefont {C.}~\bibnamefont {Cox}}, \ and\ \bibinfo
  {author} {\bibfnamefont {R.}~\bibnamefont {Feres}},\ }\href@noop {}
  {\bibfield  {journal} {\bibinfo  {journal} {Journal of Geometric Mechanics}\
  }\textbf {\bibinfo {volume} {12}},\ \bibinfo {pages} {53} (\bibinfo {year}
  {2020})}\BibitemShut {NoStop}%
\bibitem [{\citenamefont {Chernov}\ and\ \citenamefont
  {Markarian}(2006)}]{chernov}%
  \BibitemOpen
  \bibfield  {author} {\bibinfo {author} {\bibfnamefont {N.}~\bibnamefont
  {Chernov}}\ and\ \bibinfo {author} {\bibfnamefont {R.}~\bibnamefont
  {Markarian}},\ }\href {\doibase 10.1090/surv/127} {\emph {\bibinfo {title}
  {Chaotic billiards}}},\ Vol.\ \bibinfo {volume} {127}\ (\bibinfo  {publisher}
  {American Mathematical Soc.},\ \bibinfo {year} {2006})\ pp.\ \bibinfo {pages}
  {xii+316}\BibitemShut {NoStop}%
\bibitem [{Note3()}]{Note3}%
  \BibitemOpen
  \bibinfo {note} {An interesting exception to this, however, occurs in the
  case of a particle which is half rough and half smooth.\cite
  {CF}}\BibitemShut {NoStop}%
\bibitem [{\citenamefont {Cox}\ \emph {et~al.}(2018)\citenamefont {Cox},
  \citenamefont {Feres},\ and\ \citenamefont {Zhang}}]{CFZ}%
  \BibitemOpen
  \bibfield  {author} {\bibinfo {author} {\bibfnamefont {C.}~\bibnamefont
  {Cox}}, \bibinfo {author} {\bibfnamefont {R.}~\bibnamefont {Feres}}, \ and\
  \bibinfo {author} {\bibfnamefont {H.-K.}\ \bibnamefont {Zhang}},\ }\href
  {\doibase 10.1088/1361-6544/aacc43} {\bibfield  {journal} {\bibinfo
  {journal} {Nonlinearity}\ }\textbf {\bibinfo {volume} {31}},\ \bibinfo
  {pages} {4443} (\bibinfo {year} {2018})}\BibitemShut {NoStop}%
\bibitem [{\citenamefont {Wojtkowski}(1994)}]{W}%
  \BibitemOpen
  \bibfield  {author} {\bibinfo {author} {\bibfnamefont {M.~P.}\ \bibnamefont
  {Wojtkowski}},\ }\href {\doibase 10.1016/0167-2789(94)90009-4} {\bibfield
  {journal} {\bibinfo  {journal} {Phys. D}\ }\textbf {\bibinfo {volume} {71}},\
  \bibinfo {pages} {430} (\bibinfo {year} {1994})}\BibitemShut {NoStop}%
\bibitem [{\citenamefont {Kaloshin}\ and\ \citenamefont
  {Sorrentino}(2018)}]{KS}%
  \BibitemOpen
  \bibfield  {author} {\bibinfo {author} {\bibfnamefont {V.}~\bibnamefont
  {Kaloshin}}\ and\ \bibinfo {author} {\bibfnamefont {A.}~\bibnamefont
  {Sorrentino}},\ }\href
  {https://www.jstor.org/stable/10.4007/annals.2018.188.1.6} {\bibfield
  {journal} {\bibinfo  {journal} {Annals of Mathematics}\ }\textbf {\bibinfo
  {volume} {188}},\ \bibinfo {pages} {315} (\bibinfo {year}
  {2018})}\BibitemShut {NoStop}%
\bibitem [{\citenamefont {Borisov}\ \emph
  {et~al.}(2011{\natexlab{b}})\citenamefont {Borisov}, \citenamefont {Kilin},\
  and\ \citenamefont {Mamaev}}]{nonholonomic}%
  \BibitemOpen
  \bibfield  {author} {\bibinfo {author} {\bibfnamefont {A.~V.}\ \bibnamefont
  {Borisov}}, \bibinfo {author} {\bibfnamefont {A.~A.}\ \bibnamefont {Kilin}},
  \ and\ \bibinfo {author} {\bibfnamefont {I.~S.}\ \bibnamefont {Mamaev}},\
  }\href {\doibase 10.1134/S1560354711060062} {\bibfield  {journal} {\bibinfo
  {journal} {Regular and Chaotic Dynamics}\ }\textbf {\bibinfo {volume} {16}}
  (\bibinfo {year} {2011}{\natexlab{b}}),\
  10.1134/S1560354711060062}\BibitemShut {NoStop}%
\bibitem [{\citenamefont {Chen}\ \emph {et~al.}(2013)\citenamefont {Chen},
  \citenamefont {Mohr}, \citenamefont {Zhang},\ and\ \citenamefont
  {Zhang}}]{lemonergo}%
  \BibitemOpen
  \bibfield  {author} {\bibinfo {author} {\bibfnamefont {J.}~\bibnamefont
  {Chen}}, \bibinfo {author} {\bibfnamefont {L.}~\bibnamefont {Mohr}}, \bibinfo
  {author} {\bibfnamefont {H.-K.}\ \bibnamefont {Zhang}}, \ and\ \bibinfo
  {author} {\bibfnamefont {P.}~\bibnamefont {Zhang}},\ }\href {\doibase
  10.1063/1.4850815} {\bibfield  {journal} {\bibinfo  {journal} {Chaos: An
  Interdisciplinary Journal of Nonlinear Science}\ }\textbf {\bibinfo {volume}
  {23}},\ \bibinfo {pages} {043137} (\bibinfo {year} {2013})},\ \Eprint
  {http://arxiv.org/abs/https://doi.org/10.1063/1.4850815}
  {https://doi.org/10.1063/1.4850815} \BibitemShut {NoStop}%
\bibitem [{\citenamefont {Correia}\ and\ \citenamefont {Zhang}(2015)}]{moon}%
  \BibitemOpen
  \bibfield  {author} {\bibinfo {author} {\bibfnamefont {M.~F.}\ \bibnamefont
  {Correia}}\ and\ \bibinfo {author} {\bibfnamefont {H.-K.}\ \bibnamefont
  {Zhang}},\ }\href {\doibase 10.1063/1.4928594} {\bibfield  {journal}
  {\bibinfo  {journal} {Chaos: An Interdisciplinary Journal of Nonlinear
  Science}\ }\textbf {\bibinfo {volume} {25}},\ \bibinfo {pages} {083110}
  (\bibinfo {year} {2015})},\ \Eprint
  {http://arxiv.org/abs/https://doi.org/10.1063/1.4928594}
  {https://doi.org/10.1063/1.4928594} \BibitemShut {NoStop}%
\bibitem [{\citenamefont {Bunimovich}\ \emph {et~al.}(2016)\citenamefont
  {Bunimovich}, \citenamefont {Zhang},\ and\ \citenamefont {Zhang}}]{BunZhang}%
  \BibitemOpen
  \bibfield  {author} {\bibinfo {author} {\bibfnamefont {L.}~\bibnamefont
  {Bunimovich}}, \bibinfo {author} {\bibfnamefont {H.-K.}\ \bibnamefont
  {Zhang}}, \ and\ \bibinfo {author} {\bibfnamefont {P.}~\bibnamefont
  {Zhang}},\ }\href {\doibase 10.1007/s00220-015-2539-x} {\bibfield  {journal}
  {\bibinfo  {journal} {Communications in Mathematical Physics}\ }\textbf
  {\bibinfo {volume} {341}},\ \bibinfo {pages} {781} (\bibinfo {year}
  {2016})}\BibitemShut {NoStop}%
\bibitem [{\citenamefont {Correia}\ \emph {et~al.}(2017)\citenamefont
  {Correia}, \citenamefont {Cox},\ and\ \citenamefont
  {Zhang}}]{umbrellaergodicity}%
  \BibitemOpen
  \bibfield  {author} {\bibinfo {author} {\bibfnamefont {M.}~\bibnamefont
  {Correia}}, \bibinfo {author} {\bibfnamefont {C.}~\bibnamefont {Cox}}, \ and\
  \bibinfo {author} {\bibfnamefont {H.-K.}\ \bibnamefont {Zhang}},\ }\href@noop
  {} {\bibfield  {journal} {\bibinfo  {journal} {New Horizons in Mathematical
  Physics}\ } (\bibinfo {year} {2017})}\BibitemShut {NoStop}%
\bibitem [{\citenamefont {Kac}(1966)}]{Kac}%
  \BibitemOpen
  \bibfield  {author} {\bibinfo {author} {\bibfnamefont {M.}~\bibnamefont
  {Kac}},\ }\href {http://www.jstor.org/stable/2313748} {\bibfield  {journal}
  {\bibinfo  {journal} {The American Mathematical Monthly}\ }\textbf {\bibinfo
  {volume} {73}},\ \bibinfo {pages} {1} (\bibinfo {year} {1966})}\BibitemShut
  {NoStop}%
\bibitem [{\citenamefont {Ivrii}(1980)}]{ivrii}%
  \BibitemOpen
  \bibfield  {author} {\bibinfo {author} {\bibfnamefont {V.~Y.}\ \bibnamefont
  {Ivrii}},\ }\href {\doibase 10.1007/BF01086550} {\bibfield  {journal}
  {\bibinfo  {journal} {Functional Analysis and Its Applications}\ }\textbf
  {\bibinfo {volume} {14}},\ \bibinfo {pages} {98} (\bibinfo {year}
  {1980})}\BibitemShut {NoStop}%
\bibitem [{\citenamefont {Fierobe}(2021)}]{fierobe2021thesis}%
  \BibitemOpen
  \bibfield  {author} {\bibinfo {author} {\bibfnamefont {C.}~\bibnamefont
  {Fierobe}},\ }\href@noop {} {\enquote {\bibinfo {title} {Thesis manuscript:
  Projective and complex billiards, periodic orbits and pfaffian systems},}\ }
  (\bibinfo {year} {2021}),\ \Eprint {http://arxiv.org/abs/2107.01888}
  {arXiv:2107.01888 [math.DS]} \BibitemShut {NoStop}%
\bibitem [{\citenamefont {Blumen}\ \emph {et~al.}(2012)\citenamefont {Blumen},
  \citenamefont {Kim}, \citenamefont {Nance},\ and\ \citenamefont
  {Zharnitsky}}]{reflectivesphere}%
  \BibitemOpen
  \bibfield  {author} {\bibinfo {author} {\bibfnamefont {V.}~\bibnamefont
  {Blumen}}, \bibinfo {author} {\bibfnamefont {K.~Y.}\ \bibnamefont {Kim}},
  \bibinfo {author} {\bibfnamefont {J.}~\bibnamefont {Nance}}, \ and\ \bibinfo
  {author} {\bibfnamefont {V.}~\bibnamefont {Zharnitsky}},\ }\href {\doibase
  10.1093/imrn/rnr228} {\bibfield  {journal} {\bibinfo  {journal}
  {International Mathematics Research Notices}\ }\textbf {\bibinfo {volume}
  {2012}},\ \bibinfo {pages} {5014} (\bibinfo {year} {2012})}\BibitemShut
  {NoStop}%
\bibitem [{\citenamefont {Fierobe}(2020)}]{fierobe2020projective}%
  \BibitemOpen
  \bibfield  {author} {\bibinfo {author} {\bibfnamefont {C.}~\bibnamefont
  {Fierobe}},\ }\href@noop {} {\enquote {\bibinfo {title} {On projective
  billiards with open subsets of triangular orbits},}\ } (\bibinfo {year}
  {2020}),\ \Eprint {http://arxiv.org/abs/2005.02012} {arXiv:2005.02012
  [math.DS]} \BibitemShut {NoStop}%
\bibitem [{\citenamefont {Albers}\ \emph {et~al.}(2019)\citenamefont {Albers},
  \citenamefont {Banhatti}, \citenamefont {Sadlo}, \citenamefont {Schwartz},\
  and\ \citenamefont {Tabachnikov}}]{albers2019polygonal}%
  \BibitemOpen
  \bibfield  {author} {\bibinfo {author} {\bibfnamefont {P.}~\bibnamefont
  {Albers}}, \bibinfo {author} {\bibfnamefont {G.}~\bibnamefont {Banhatti}},
  \bibinfo {author} {\bibfnamefont {F.}~\bibnamefont {Sadlo}}, \bibinfo
  {author} {\bibfnamefont {R.}~\bibnamefont {Schwartz}}, \ and\ \bibinfo
  {author} {\bibfnamefont {S.}~\bibnamefont {Tabachnikov}},\ }\href@noop {}
  {\enquote {\bibinfo {title} {Polygonal symplectic billiards},}\ } (\bibinfo
  {year} {2019}),\ \Eprint {http://arxiv.org/abs/1912.09404} {arXiv:1912.09404
  [math.SG]} \BibitemShut {NoStop}%
\end{thebibliography}%

\end{document}